\documentclass[11pt]{article}

\usepackage{moriond,epsfig}

\newcommand\aj{AJ}%                                         % Astronomical Journal
\newcommand\apj{ApJ}%                                       % Astrophysical Journal
\newcommand\apjs{ApJS}%                                     % Astrophysical Journal, Supplement
\newcommand\aap{A$\&$A}%                                    % Astronomy and Astrophysics
\newcommand\mnras{MNRAS}%                                   % Monthly Notices of the RAS
\newcommand\prd{Phys.~Rev.~D}%                              % Physical Review D
\newcommand\physrep{Phys.~Rep.}%                            % Physics Reports
\def\be{\begin{equation}}
\def\ee{\end{equation}}
\def\bea{\begin{eqnarray}}
\def\eea{\end{eqnarray}}

\begin{document}

%%%%%%%%%%%%%%%%%%%%%%%%%%%%%%%%%%%%%%%%%%%%%%%%%%%%%%%%%%%%%%%%%%%%%%%%%%%%%%%%%%%%%%%%%%%%%%%%%%%%%%%%%%%%%%%%%%%%%%%%%%%%%%%%%%%%%%%%%%%%%%%%%%%%%%%%%%%%%%%%%

% TITLE

\vspace*{4cm}
\title{	NEUTRINO MASS FROM THE LYMAN-ALPHA FOREST}

\author{GRAZIANO ROSSI}

\address{Department of Astronomy and Space Science, Sejong University, Seoul, 143-747, Korea}

%%%%%%%%%%%%%%%%%%%%%%%%%%%%%%%%%%%%%%%%%%%%%%%%%%%%%%%%%%%%%%%%%%%%%%%%%%%%%%%%%%%%%%%%%%%%%%%%%%%%%%%%%%%%%%%%%%%%%%%%%%%%%%%%%%%%%%%%%%%%%%%%%%%%%%%%%%%%%%%%%

% ABSTRACT

\maketitle\abstracts{
The quest for the neutrino mass is a central goal
in contemporary cosmology, subject to intense scrutiny,
and among different large-scale structure tracers the Lyman-$\alpha$ forest is re-emerging as a unique 
tool to probe the neutrino mass at high-redshift -- through characteristic imprints on the
transmitted Lyman-$\alpha$ flux.
A detailed modeling of the low-density regions of the intergalactic medium in presence of massive neutrinos
on scale ranging from a few to hundreds of megaparsecs is required, if 
one wants to interpret state-of-the-art data from observations of quasar spectra.
To this end, we provide a suite of hydrodynamical simulations made with Gagdet-3, spanning different volumes and having a range of resolutions and
neutrino masses (from $M_{\rm \nu}=0.1$ to $0.8$ eV, assuming 3 degenerate species), specifically designed to meet the
requirements of the Baryon Acoustic Spectroscopic Survey (BOSS).
We adopted a particle-type implementation of massive neutrinos, and chose
cosmological parameters compatible with the latest Planck (2013) results.
While the resolution requirements match the quality of the SDSS-III/BOSS
data, our numerical simulations will also establish a useful theoretical ground for upcoming surveys such as SDSS-IV/eBOSS and DESI.
In the very near future, data from leading spectroscopic surveys will allow measuring the absolute mass scale of neutrinos, and
determining the exact nature of the neutrino mass hierarchy; hence, we expect that this modeling will become increasingly useful.
}

%%%%%%%%%%%%%%%%%%%%%%%%%%%%%%%%%%%%%%%%%%%%%%%%%%%%%%%%%%%%%%%%%%%%%%%%%%%%%%%%%%%%%%%%%%%%%%%%%%%%%%%%%%%%%%%%%%%%%%%%%%%%%%%%%%%%%%%%%%%%%%%%%%%%%%%%%%%%%%%%%

% MASSIVE NEUTRINOS AND STRUCTURE FORMATION

\section{Massive Neutrinos and Structure Formation}

%-----------------------------------------------------------------------------------

\begin{figure}[t]
\begin{center}
\psfig{figure=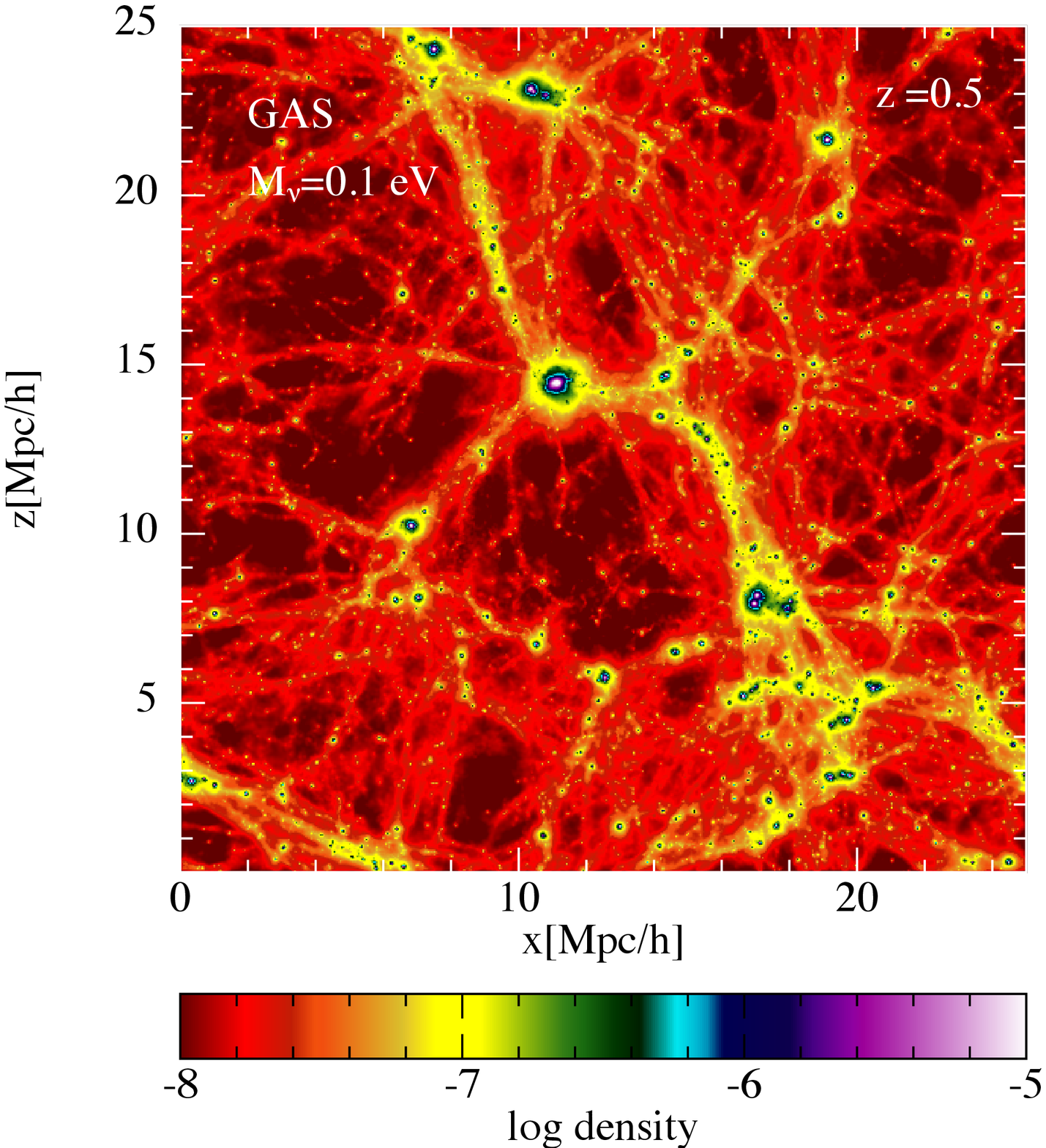,height=2.065in}
\psfig{figure=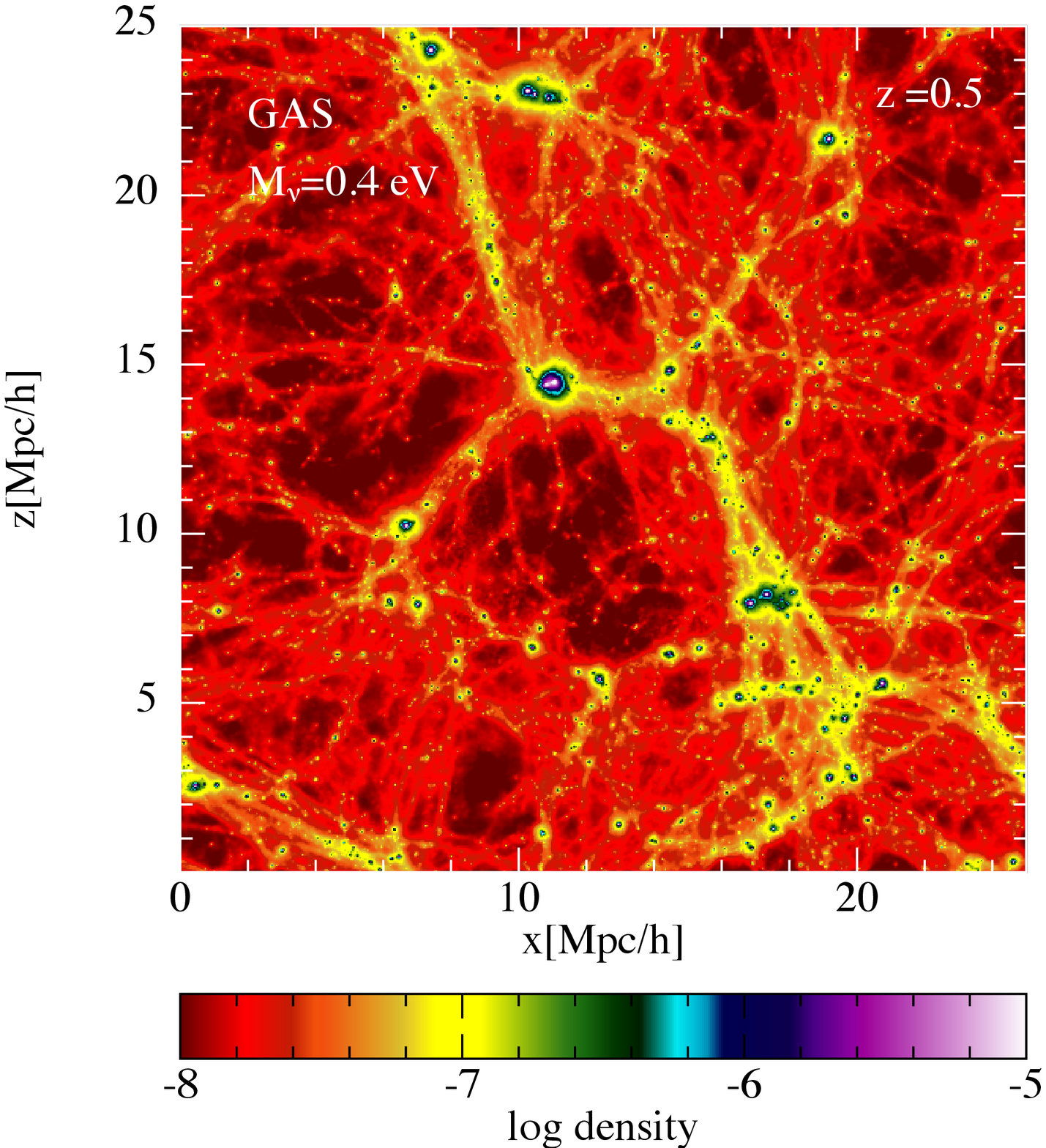,height=2.065in}
\psfig{figure=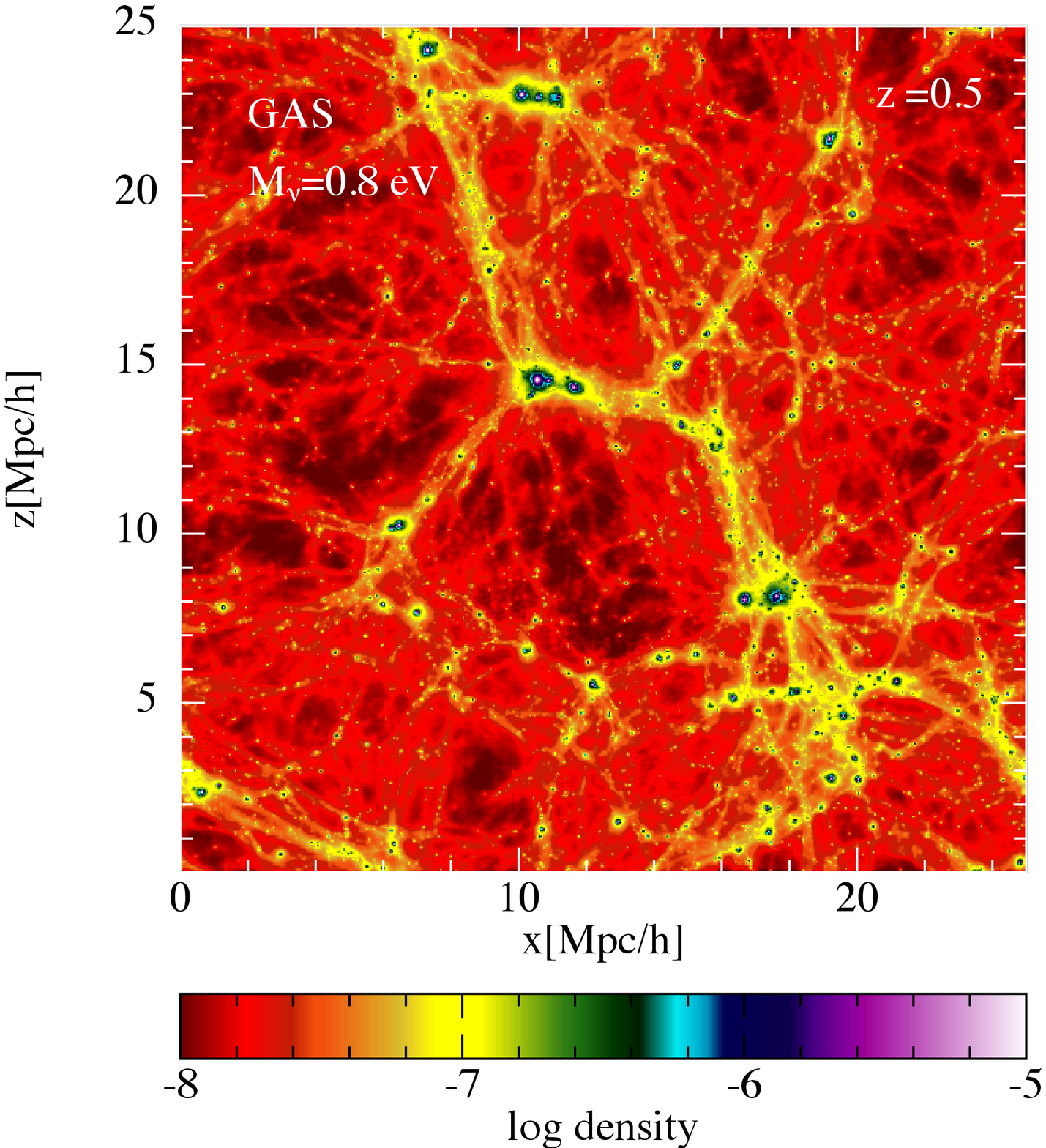,height=2.065in}
\psfig{figure=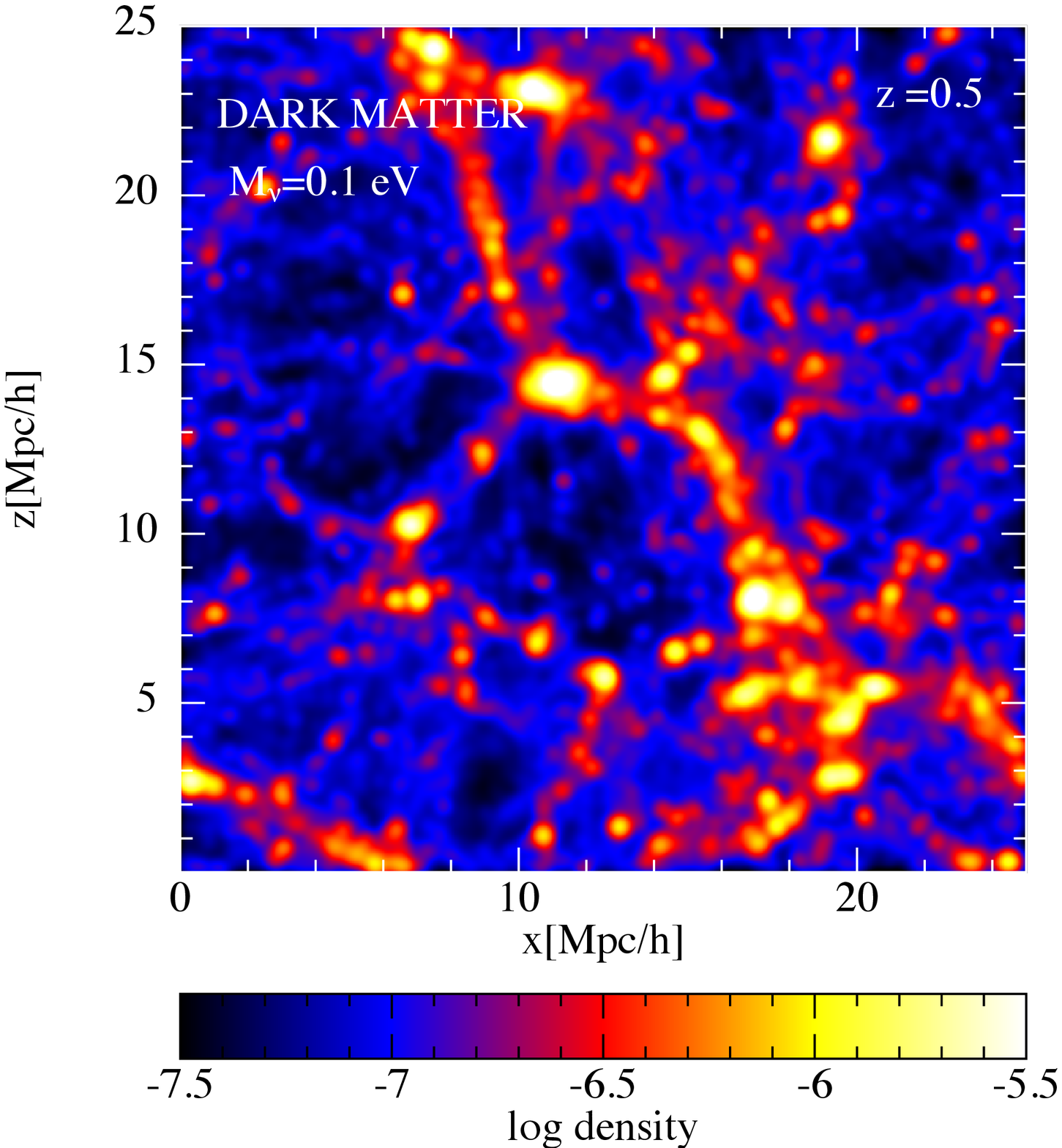,height=2.065in}
\psfig{figure=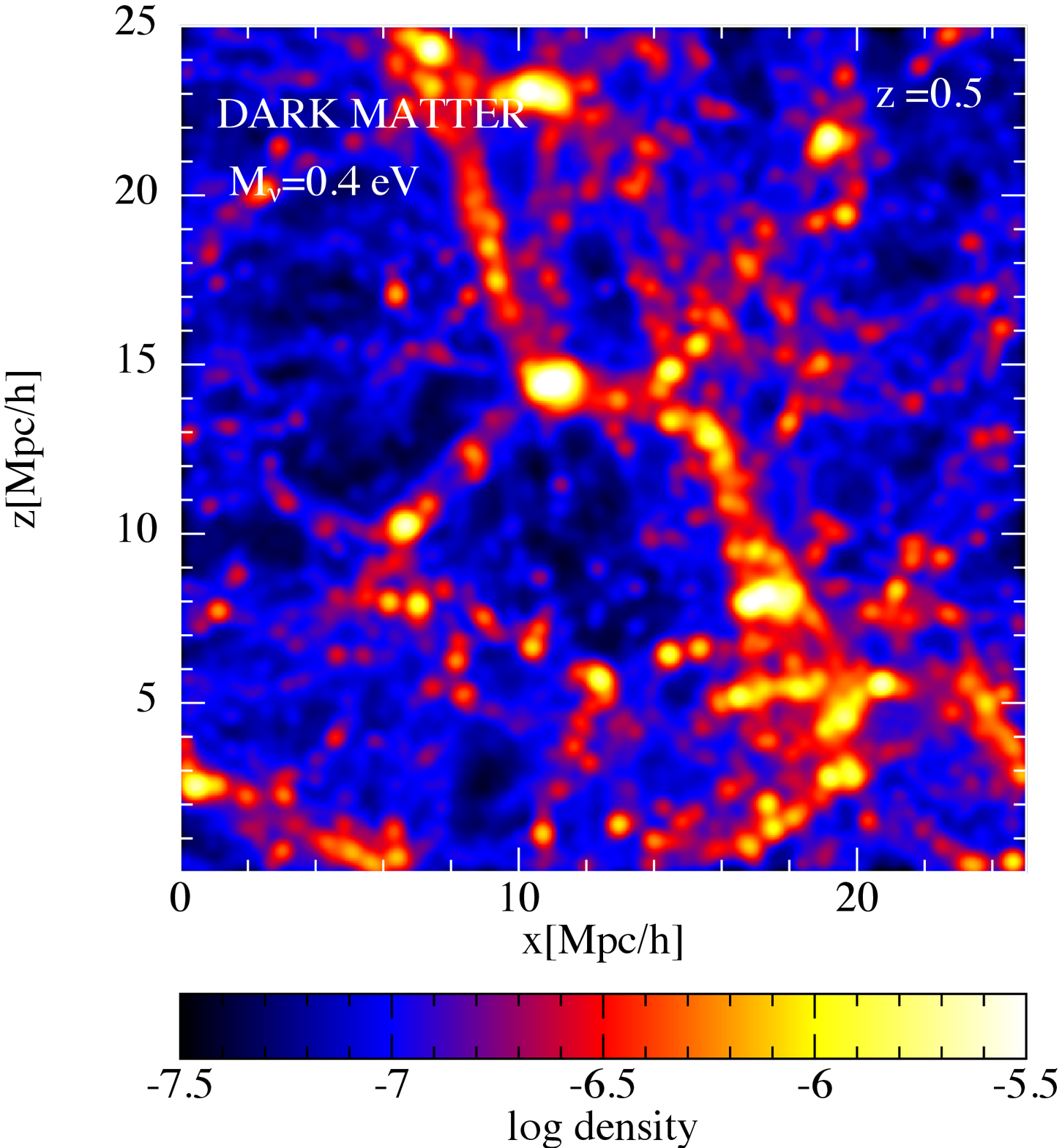,height=2.065in}
\psfig{figure=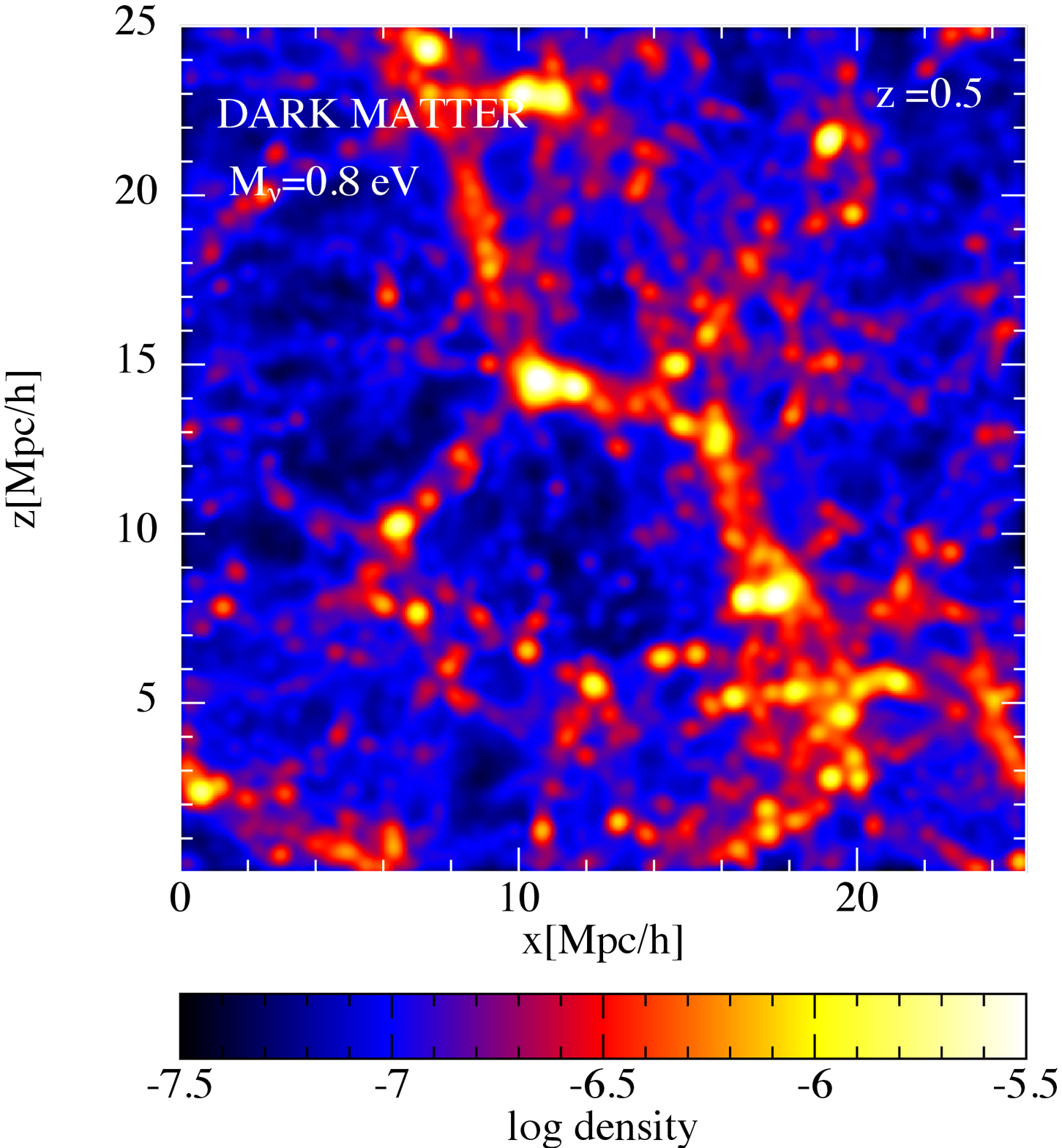,height=2.065in}
\caption{Examples of snapshots of the gas (top panels) and dark matter (bottom panels) components at $z=0.5$,
from simulations with 25 $h^{-1}$Mpc box size and resolution $N_{\rm p}=192^3$ particles/type.
The panels are full projections of the density field in the $x$ and $z$ directions across $y$ and smoothed with a cubic spline kernel, 
obtained from simulations having a 
total neutrino mass $M_{\rm \nu}=0.1$ eV (left), $M_{\rm \nu}=0.4$ eV (middle), and $M_{\rm \nu}=0.8$ eV (right).}
\label{fig1}
\end{center}
\end{figure}

%-----------------------------------------------------------------------------------

% General Intro: Neutrinos in Cosmology

At the interface between particle physics and cosmology,
neutrino science has received renewed interest
recently, after the breakthrough discovery over the last decade that neutrinos are massive.
Hence, massive neutrinos should be included 
in the concordance $\Lambda$CDM  model 
dominated by a dark energy (DE) component, which in general
only assumes a minimal neutrino mass of 0.06 eV.
While neutrino oscillation experiments are sensitive only to differences in the squares of neutrino masses,
 cosmology offers a unique `laboratory' with the best sensitivity to the neutrino mass: 
 primordial neutrinos leave their imprint into several large-scale structure (LSS) observables, and because of free-streaming they
 significantly alter structure formation.  Therefore, 
cosmology can place competitive limits on the neutrino mass scale and hierarchy.
 
 % A Bit More on Neutrino Science
 
 Neutrinos must be considered as extra radiation while ultra-relativistic, but
 once non-relativistic they behave as an additional cold dark matter (CDM) component and 
 participate in structure formation
 on scales greater than their free-streaming scale: the overall result is a  
 suppression of power on small scales, and a 
 delay in matter domination (Lesgourgues \& Pastor 2006).
 Neutrinos in the mass range 0.05 eV $\le m_{\rm \nu} \le$ 1.5 eV become non-relativistic
 in the redshift interval $3000 \ge z \ge 100$, approximately around $z_{\rm nr} \simeq 2000 (m_{\rm \nu}/1{\rm eV})$.
While the effect of cosmological neutrinos on the evolution of density perturbations in the linear regime is well-understood, 
 less is known in the nonlinear regime: this fact motivates the present study.
 
 % Different Probes and the Lyman-Alpha Forest
 
 Massive neutrinos can be studied through their impact on
 the CMB, particularly in the polarization maps, and by using
 several baryonic tracers of the LSS clustering of matter such as
 the 3D power spectrum from galaxy surveys,
 the Sunyaev-Zel'dovich effect in galaxy clusters, cosmic shear
 through weak lensing, or the Lyman-$\alpha$ (Ly$\alpha$) forest.
 The latter observable has generally received less attention in the literature, 
but is currently emerging as a promising window
 into the high-redshift Universe, being at a redshift range inaccessible to
 other LSS probes and spanning a wide interval in redshift. In particular, 
 the suppression of growth of cosmological structures
 on scales smaller than the neutrino free-streaming distance
 makes the Ly$\alpha$ forest a good tracer of the neutrino mass,
 and measurements of the mean Ly$\alpha$
 transmission flux and its evolution allow constraining the basic cosmological parameters
 with improved sensitivity.
 
 % The BOSS Survey and Exquisite Unique Data, and Simulations
 
At the present time, the best Ly$\alpha$ forest data
and the most  precise measurement of the Ly$\alpha$
flux power spectrum come from the Baryon Acoustic Spectroscopic Survey (BOSS) -- see Dawson et al. (2013), 
and Palanque-Delabrouille et al. (2013). In order to interpret the information contained in this remarkable data set and control the systematics involved,  
numerical simulations at equivalent or superior precision of this survey are required,
particularly at lower redshifts and smaller scales
 (1--40 $h^{-1}$Mpc) where for massive neutrinos the nonlinear evolution of
 density fluctuations becomes significant.
Despite their  intrinsic limitations and uncertainties,
simulations allow one to self-consistently
 model the interplay between gravity and gas pressure on the structure of the
 photoionized intergalactic medium (IGM), so that most of the observed properties of the Ly$\alpha$ forest are 
 reproduced, and to gain a better understanding of the role and effects of massive neutrinos in the complex 
 process of structure formation -- as we discuss next. 
 
%%%%%%%%%%%%%%%%%%%%%%%%%%%%%%%%%%%%%%%%%%%%%%%%%%%%%%%%%%%%%%%%%%%%%%%%%%%%%%%%%%%%%%%%%%%%%%%%%%%%%%%%%%%%%%%%%%%%%%%%%%%%%%%%%%%%%%%%%%%%%%%%%%%%%%%%%%%%%%%%%

% SUITE OF HYDRODYNAMICAL SIMULATIONS WITH MASSIVE NEUTRINOS

\section{Suite of Hydrodynamical Simulations with Massive Neutrinos}

%-----------------------------------------------------------------------------------

\begin{figure}[t]
\begin{center}
\psfig{figure=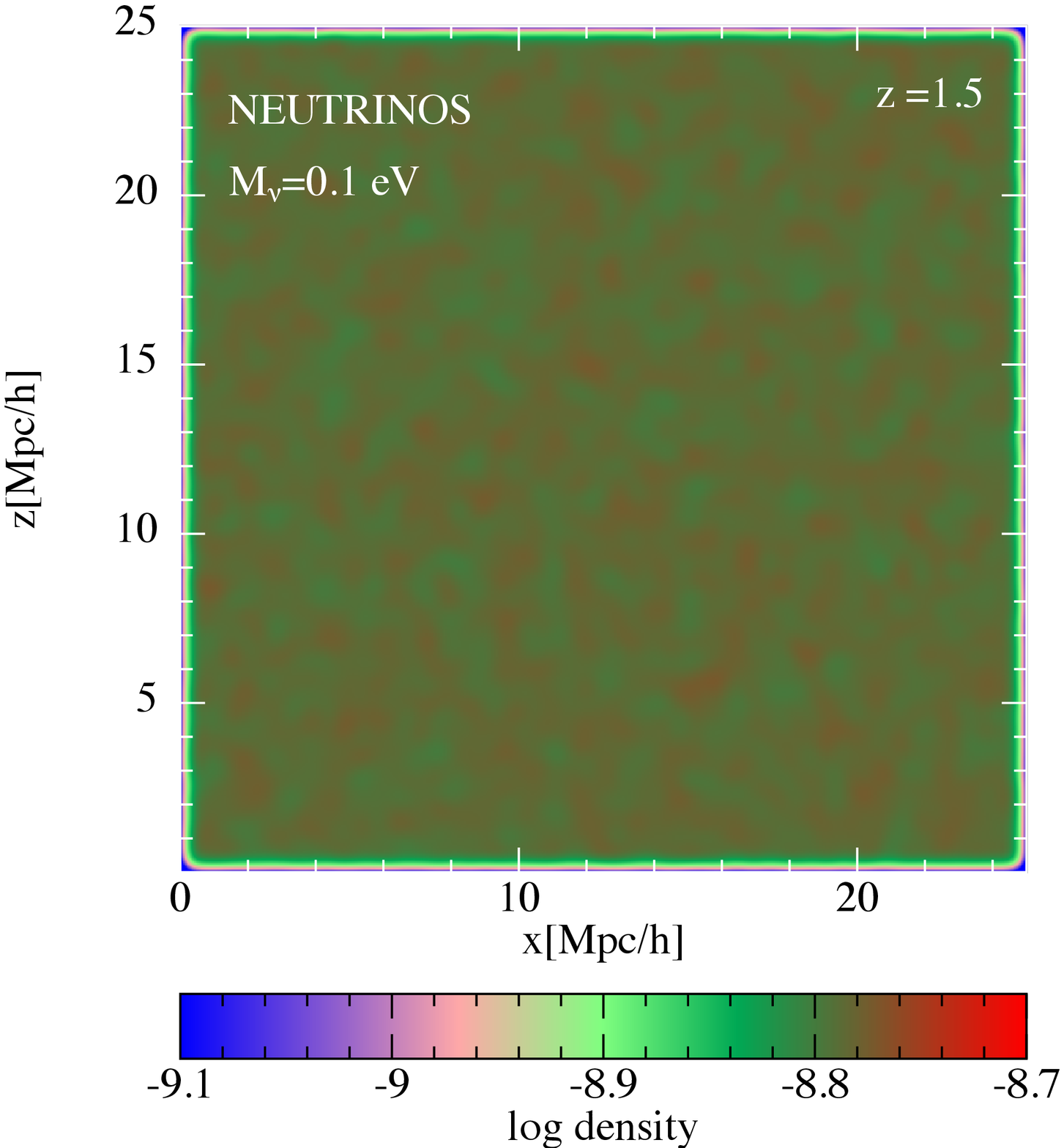,height=2.065in}
\psfig{figure=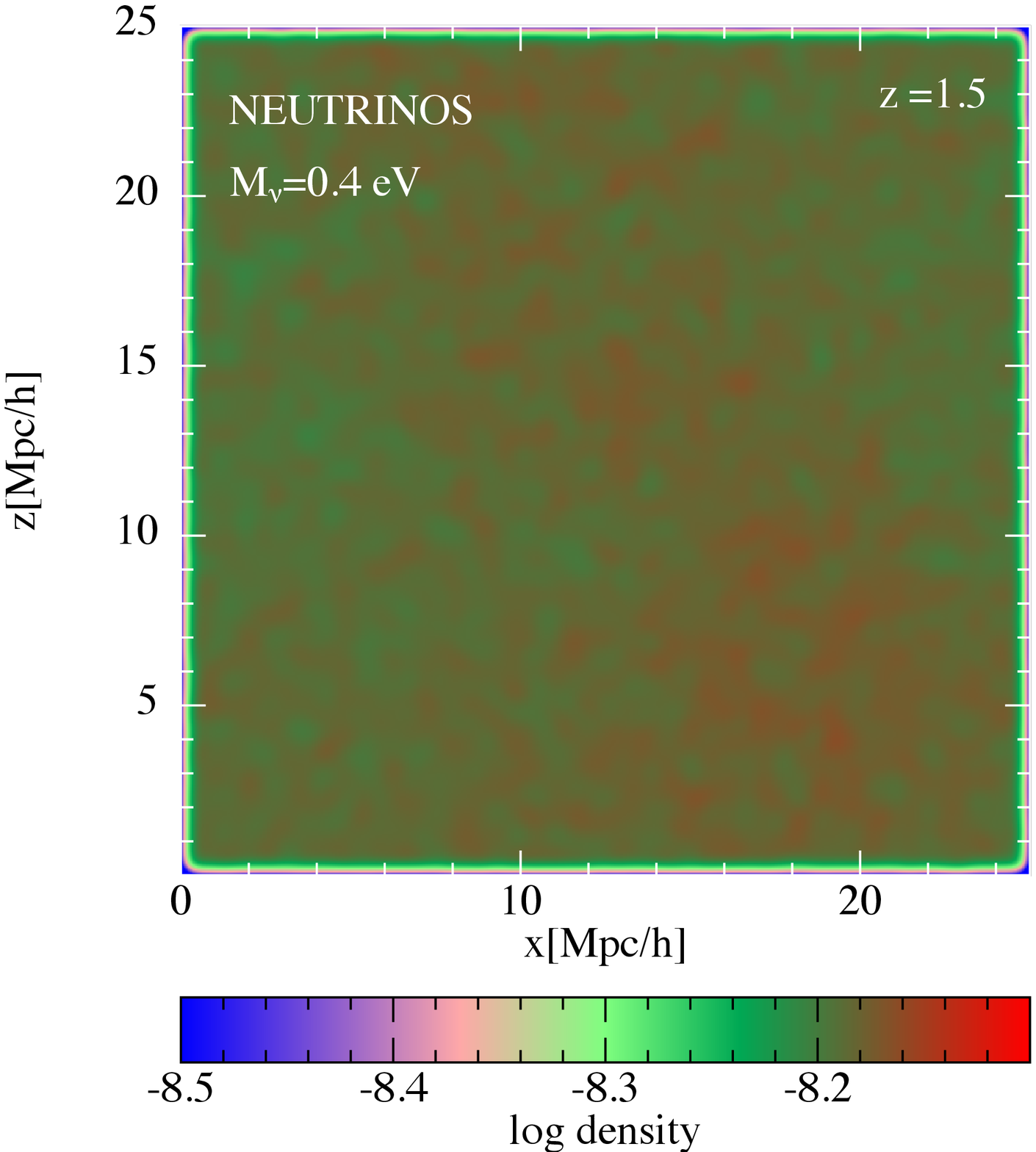,height=2.065in}
\psfig{figure=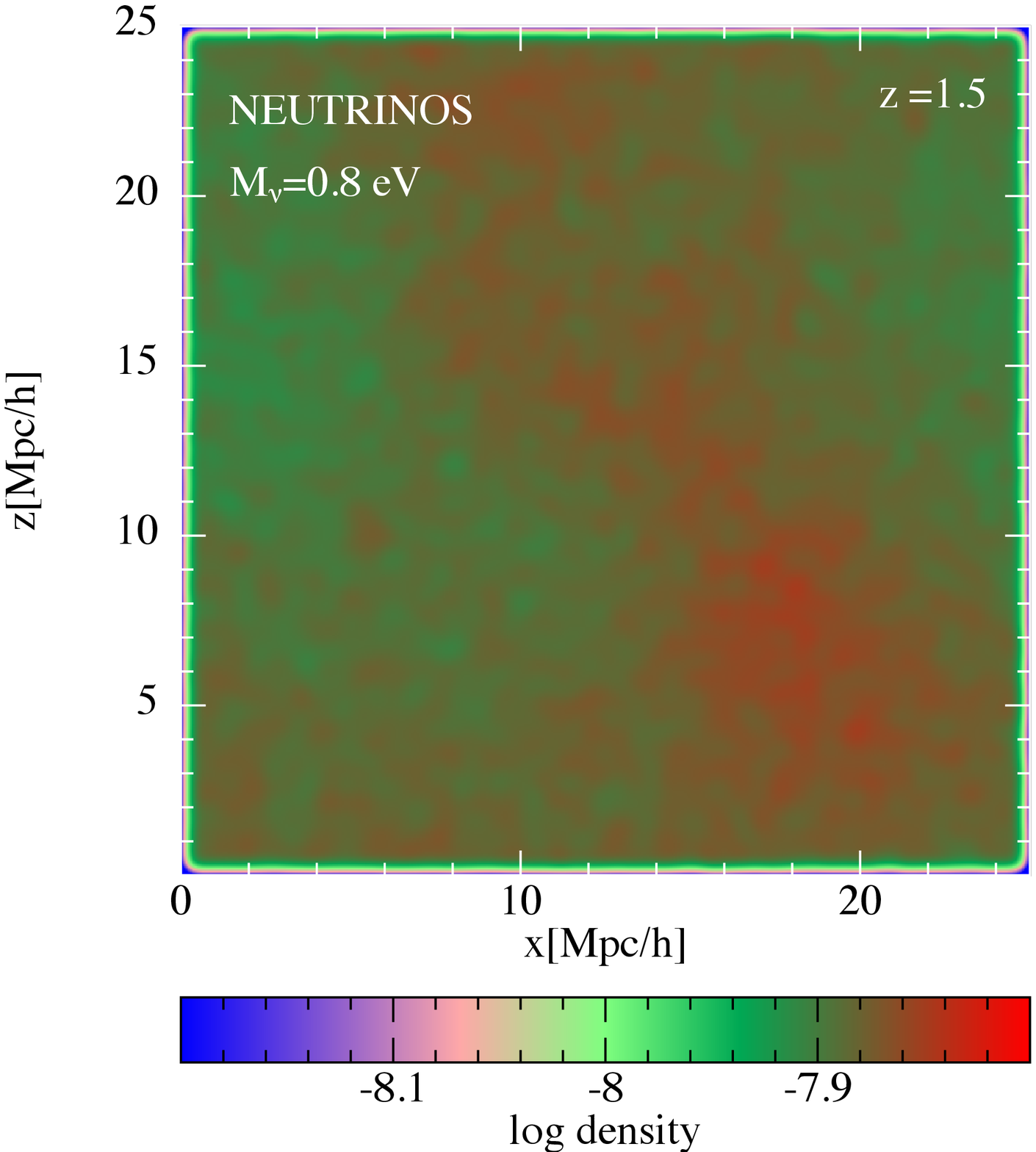,height=2.065in}
\psfig{figure=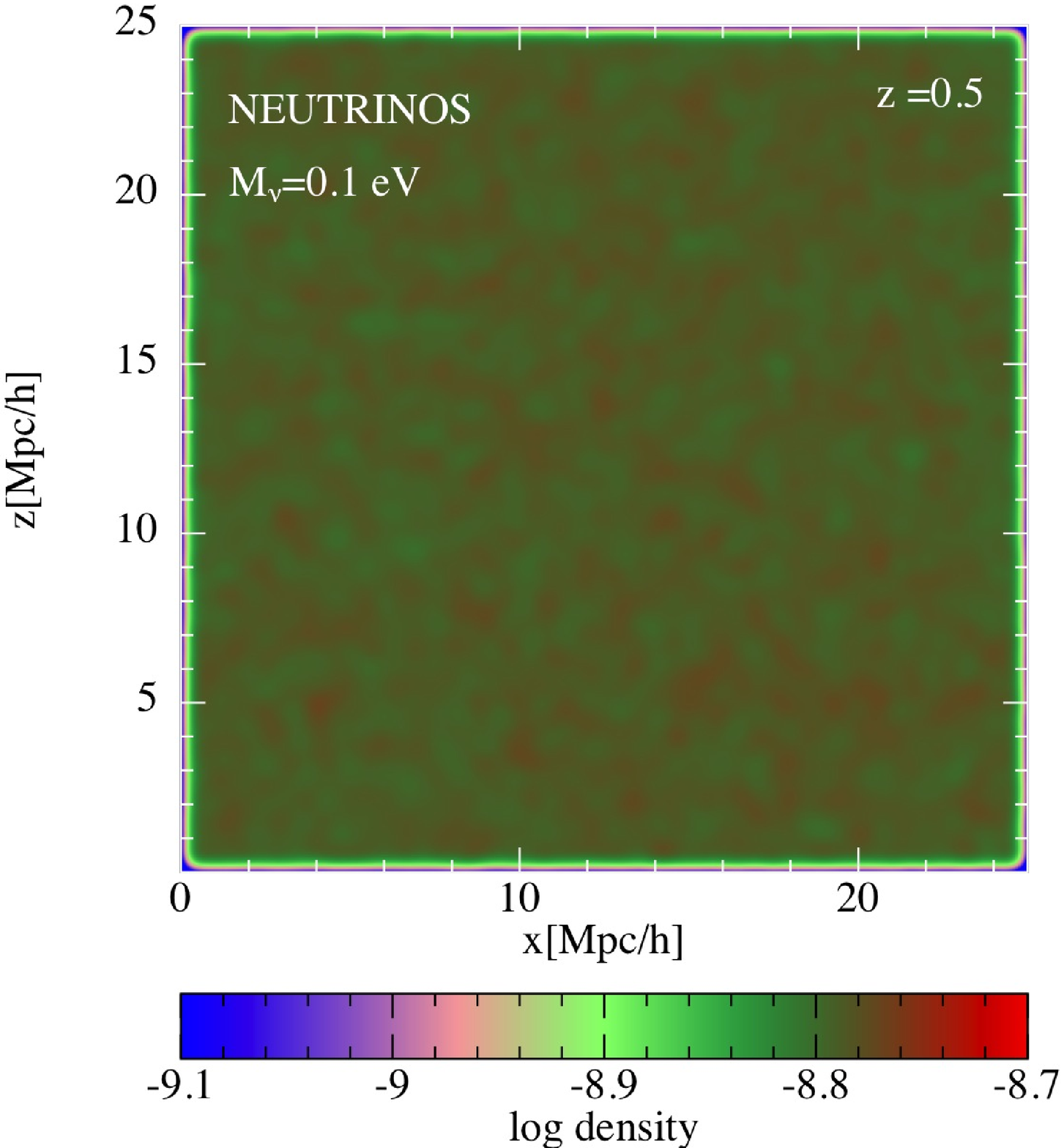,height=2.065in}
\psfig{figure=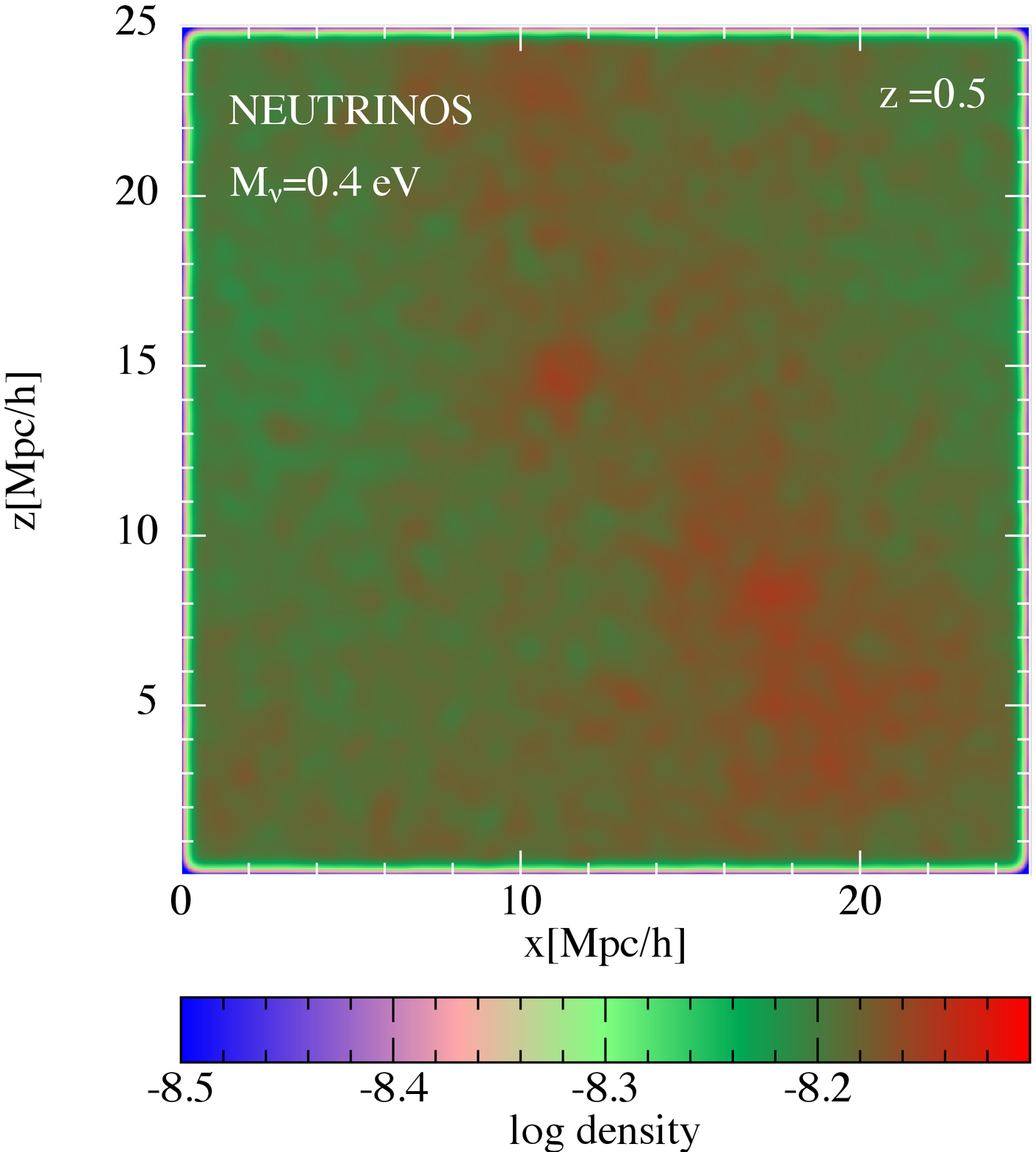,height=2.065in}
\psfig{figure=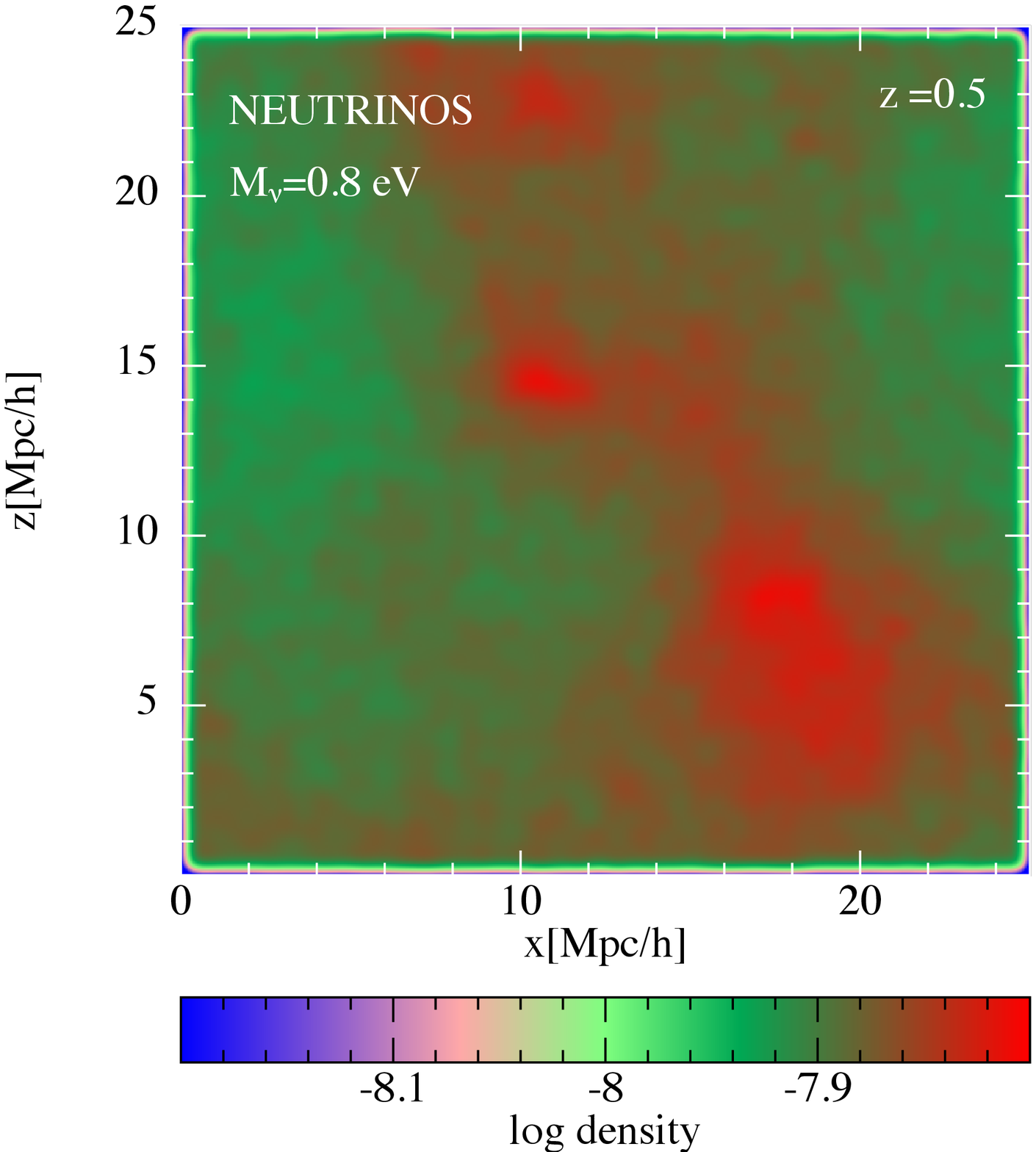,height=2.065in}
\caption{Density evolution of the neutrino component at $z=1.5$ (top panels) and $z=0.5$ (bottom panels),
from simulations with $25~h^{-1}$Mpc box size and resolution $N_{\rm p}=192^3$/type.
The total neutrino mass increases from left to right, being $M_{\rm \nu}=0.1$ eV (left), $M_{\rm \nu}=0.4$ eV (central), and $M_{\rm \nu}=0.8$ eV (right). 
The distribution of the neutrino density has been smoothed with a cubic spline
kernel to eliminate spurious Poisson noise at small scales.}
\label{fig2}
\end{center}
\end{figure}

%-----------------------------------------------------------------------------------

We have produced a suite 
of 48 cosmological hydrodynamical simulations
with CDM, 
baryons, and either a  varying neutrino mass
and fixed cosmological and astrophysical parameters,
or with a fixed neutrino mass and slight variations in the basic cosmological and astrophysical parameters
around what we termed the `best-guess' -- namely, the reference simulation set with 
Planck (2013) cosmological parameters and a massless neutrino component.
Box sizes and resolutions range from  25 $h^{-1}$Mpc to 100 $h^{-1}$Mpc, and 
from  $N_{\rm p}=192^3$ to  $N_{\rm p}=768^3$ particles/type, respectively.
Visual examples of the density distribution of the 
gas and dark matter components at $z=0.5$
in cosmologies with massive neutrinos are shown in Figure \ref{fig1}.
Extensive details  on the numerical aspects of these simulations, and on the implementation of massive neutrinos,
 can be found in Rossi et al. (2014).
In particular, along the lines of 
Viel et al. (2010), we have modeled neutrinos as
an additional type of particle
in the $N$-body setup, and carried out a full hydrodynamical treatment well-inside
the nonlinear regime, without making any approximations for the evolution of the neutrino component.
We have considered 3 degenerate species of massive neutrinos implemented as a 
single particle-type, with total mass $M_{\rm \nu}=0.1,0.2,0.3,0.4,0.8$ eV.
Figure \ref{fig2} is a visual example of the density evolution of the neutrino component for different $M_{\rm \nu}$ ranges, at
$z=1.5$ (top panels) and $z=0.5$ (bottom panels).
Simulations were made with Gadget-3 (Springel 2005), CAMB (Lewis, Challinor \& Lasenby 2000),
and 2LPT (Crocce et al. 2006) initial conditions starting at $z=30$, and
contain improvements at all levels with respect to previous work.
We stored snapshots at redshifts between $z=4.6$ and $z=2.2$ in $\Delta z=0.2$ intervals, and for each simulation
we extracted 100,000 random pencil beam lines of sight (LOS).

%%%%%%%%%%%%%%%%%%%%%%%%%%%%%%%%%%%%%%%%%%%%%%%%%%%%%%%%%%%%%%%%%%%%%%%%%%%%%%%%%%%%%%%%%%%%%%%%%%%%%%%%%%%%%%%%%%%%%%%%%%%%%%%%%%%%%%%%%%%%%%%%%%%%%%%%%%%%%%%%%

% RESULTS, APPLICATIONS, FUTURE PROSPECTS

\section{Results, Applications, and Future Prospects}

%-----------------------------------------------------------------------------------

\begin{figure}[t]
\begin{center}
\psfig{figure=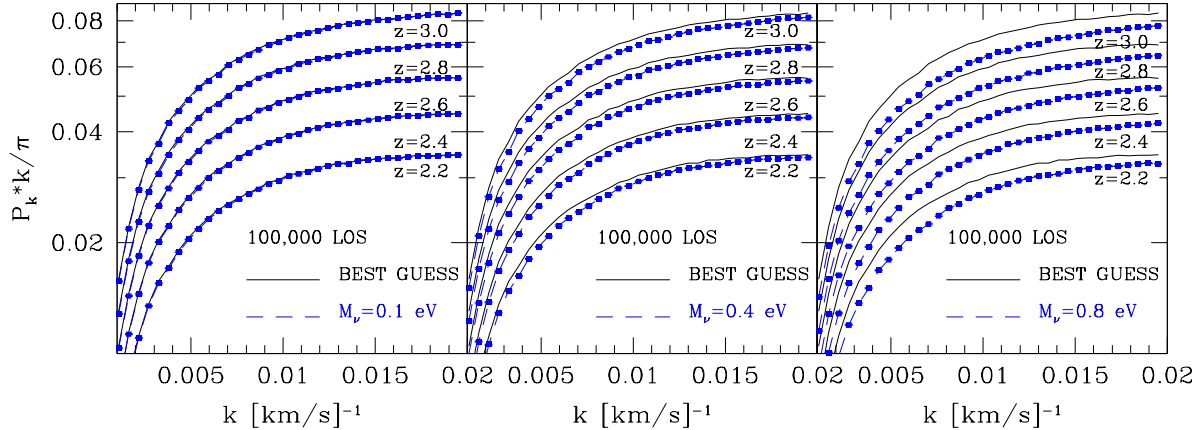,height=2.33in}
\caption{One-dimensional flux power spectra computed from the suite of simulations described in the main text, for different values of the neutrino mass (dashed lines and points)
as specified in the panels, averaged over 100,000 LOS at different $z$-intervals.
Black lines are the corresponding measurements obtained from the `best-guess', which contains only a massless neutrino component.
Results are obtained by applying the splicing technique.}
\label{fig3}
\end{center}
\end{figure}

%-----------------------------------------------------------------------------------

The free-streaming of neutrinos causes a suppression of the power spectrum of the total
matter distribution at scales probed by the Ly$\alpha$ forest data, which
is larger than the linear theory prediction by about $\sim 5\%$ at scales $k \sim 1~h$Mpc$^{-1}$
when $M_{\rm \nu}=0.4$ eV, and is strongly redshift dependent.
This effect propagates into the 1D flux power spectrum, 
and affects the statistical properties of the Ly$\alpha$ transmitted flux fraction.
Figure \ref{fig3} shows examples of 1D flux power spectra across several redshift slices, and
for a varying total neutrino mass -- i.e. $M_{\rm \nu}=0.1$ eV (left), $M_{\rm \nu}=0.4$ eV (center), $M_{\rm \nu}=0.8$ eV (right).
Results are averaged over 100,000 LOS at different $z$, as indicated in the panels, and are
obtained with 
the splicing technique introduced by McDonald (2003) and extensively tested in Borde et al. (2014).
This allows us to achieve an equivalent resolution of $3 \times 3072^3\simeq 87$ billion particles in a $(100~h^{-1}{\rm Mpc})^3$
box size with a $2\%$  global error across the full $k$-range of interest -- which is at the same level
of the current uncertainties in available observational data -- 
without the need of running a single but computationally prohibitive numerical simulation.
The flux power spectrum is sensitive to a wide range of cosmological and
astrophysical parameters and instrumental effects; it can be used as
a probe of the primordial matter power spectrum on scales of
$0.5 - 40~h^{-1}$Mpc at $2 \le z \le 4$, and 
 to determine cosmological parameters, the
nature of dark matter through its shape and redshift dependence, and the neutrino mass.

We are currently combining results of these simulations with Ly$\alpha$ forest data 
from BOSS, in order to constrain cosmological parameters and the 
neutrino mass with improved sensitivity, and fully exploit the orthogonality of the Ly$\alpha$
forest with other LSS probes. Our simulations and techniques
can also be useful for upcoming surveys such as SDSS-IV/eBOSS (Comparat et al. 2013)  and DESI (Schlegel et al. 2011), which will eventually lead to
the determination of the absolute mass scale and hierarchy of neutrinos in the very near future.

%%%%%%%%%%%%%%%%%%%%%%%%%%%%%%%%%%%%%%%%%%%%%%%%%%%%%%%%%%%%%%%%%%%%%%%%%%%%%%%%%%%%%%%%%%%%%%%%%%%%%%%%%%%%%%%%%%%%%%%%%%%%%%%%%%%%%%%%%%%%%%%%%%%%%%%%%%%%%%%%%

% ACKNOWLEDGMENTS

\section*{Acknowledgments}

This work and the participation to the `Rencontres de Moriond' (March 2014) in La Thuile, Aosta Valley, Italy, were supported by the faculty research fund of Sejong University in 2014.
It is a pleasure to
thank Jacques Dumarchez for the superb organization, along with 
the scientific and organizing committees and the secretariat. It has been an enjoyable and stimulating meeting.

%%%%%%%%%%%%%%%%%%%%%%%%%%%%%%%%%%%%%%%%%%%%%%%%%%%%%%%%%%%%%%%%%%%%%%%%%%%%%%%%%%%%%%%%%%%%%%%%%%%%%%%%%%%%%%%%%%%%%%%%%%%%%%%%%%%%%%%%%%%%%%%%%%%%%%%%%%%%%%%%%

% BIBLIOGRAPHY

\section*{Selected References}

%%%%%%%%%%%%%%%%%%%%%%%%%%%%%%%%%%%%%%%%%%%%%%%%%%%%%%%%%%%%%%%%%%%%%%%%%%%%%%%%%%%%%%%%%%%%%%%%%%%%%%%%%%%%%%%%%%%%%%%%%%%%%%%%%%%%%%%%%%%%%%%%%%%%%%%%%%%%%%%%%

\end{document}